# Unconventional isotope effect on transition temperature in BiS$_2$-based superconductor Bi$_4$O$_4$S$_3$

Rajveer Jha[1] and Yoshikazu Mizuguchi[1]*

1. Department of Physics, Tokyo Metropolitan University, 1-1, Minami-osawa, 192-0397 Hachioji, Japan.

Corresponding author: Yoshikazu Mizuguchi (mizugu@tmu.ac.jp)

**Abstract**

We report the sulfur isotope effect on transition temperature in a BiS$_2$-based superconductor Bi$_4$O$_4$S$_3$. Polycrystalline samples of Bi$_4$O$_4$S$_3$ were prepared using $^{32}$S and $^{34}$S isotope chemicals. From magnetization analyses, the isotope exponent ($\alpha_S$) was estimated as $-0.1 < \alpha_S < 0.1$. Although the $T_c$ estimated from electrical resistivity was scattered as compared to those estimated from the magnetization, we observed no clear correlation between $T_c$ and the isotope mass. The present results suggest that unconventional paring states are essential in Bi$_4$O$_4$S$_3$.

# 1. Introduction

Superconducting pairing mechanisms in most superconductors are achieved by electron-phonon interactions and explained by the Bardeen–Cooper–Schrieffer (BCS) theory [1]. However, there are unconventional superconductors whose pairing is not mediated by phonon. The typical examples of unconventional superconductors are the layered cuprate and Fe-based high-transition-temperature (high-$T_c$) superconductors [2,3], in which pairing is mediated by spin and/or orbital fluctuations [4,5]. Although there are various ways to test if the superconductivity mechanisms are mediated by phonon or other fluctuations, studying isotope effect on $T_c$ is a direct probe to examine the importance of phonon to the superconductivity. Based on the conventional phonon-mediated model, $T_c$ should be proportional to its phonon energy $\hbar\omega$, where $\hbar$ and $\omega$ are the Planck constant and phonon frequency, respectively. The $T_c$ of a conventional superconductor is, therefore, sensitive to the isotope effect of the constituent elements. The isotope exponent $\alpha$, which is defined by $T_c \sim M^{-\alpha}$, where $M$ is the isotope mass, is expected to be $\alpha \sim 0.5$ from the conventional model [1]. For example, $\alpha$ close to 0.5 has been reported for a bismuth oxide $(Ba,K)BiO_3$ ($\alpha_O \sim 0.5$) [6], doped fullerene ($\alpha_C \sim 0.4$) [7], $MgB_2$ ($\alpha_B \sim 0.3$) [8], and borocarbides ($\alpha_B \sim 0.3$) [9]. In contrast, $\alpha$ different from 0.5 has been reported for unconventional superconductors. In the cuprate superconductor system, $\alpha_O$ deviates from 0.5 and exhibits anomalous carrier concentration dependence [10,11]. In the Fe-based superconductor, an inverse isotope effect, negative $\alpha_{Fe}$, for the same composition [12].

In this study, we focus on the $BiS_2$-based layered superconductors, which was discovered in 2012 [13-15]. Since the layered structure composed of alternate stacking of a conduction layer ($BiS_2$ layer) and an insulator layer is quite similar to that of the high-$T_c$ superconductors, elucidation of the mechanisms of superconductivity in $BiS_2$-based superconductors has been desired; however, it is still controversial [16,17]. In the early stage, theoretical and experimental studies reported the possibility of conventional electron-phonon superconductivity in the $BiS_2$-based compounds [18-21]. However, some of theoretical studies suggested the possibility of unconventional mechanisms of superconductivity in the $BiS_2$-based system [22-24]. In addition, the possibility of unconventional mechanisms was experimentally suggested from angle-resolved photoemission spectroscopy in $NdO_{0.71}F_{0.29}BiS_2$ [25] and the Se-isotope effect in $LaO_{0.6}F_{0.4}BiSSe$ ($-0.04 < \alpha_{Se} < 0.04$) [26]. Therefore, further experimental investigations on the mechanisms are desired for the $BiS_2$-based superconductor family. To investigate the S isotope effect on $T_c$, we selected $Bi_4O_4S_3$ to be a target phase of this study because its $T_c$ is relatively high ($T_c > 4.5$ K), the transition is relatively sharp, and the composition is the simplest (ternary) among the $BiS_2$-based superconductors [13, 27-29]. Here, we show that the $Bi_4O_4S_3$ superconductor samples prepared with $^{32}S$ and $^{34}S$ isotope chemicals did not show conventional-type isotope effect on $T_c$, and the estimated $\alpha_S$ was close to zero. This result proposes that the unconventional isotope effect is universal feature between LaO$_{0.6}$F$_{0.4}$BiSSe with a BiSSe-type superconducting layer and Bi$_4$O$_4$S$_3$ with a BiS$_2$-type superconducting layer.

## 2. Experimental details

Polycrystalline samples of Bi$_4$O$_4$S$_3$ were prepared by a solid-state-reaction method. Powders of Bi (99.999%) and Bi$_2$O$_3$ (99.999%) were mixed with powders of $^{32}$S (ISOFLEX: 99.99%) or $^{34}$S (ISOFLEX: 99.26%) with a nominal composition of Bi$_4$O$_4$S$_3$. The mixed powder was pelletized, sealed in an evacuated quartz tube, and heated at 540 °C for 10 h. The obtained sample was examined by laboratory X-ray diffraction (XRD) by the $\theta$-$2\theta$ method with a CuK$\alpha$ radiation on Miniflex600 (RIGAKU) equipped with a Dtex-Ultra semiconductor detector. Lattice constants were calculated by Rietveld fitting by RIETAN-FP [30]. The structural image was drawn by VESTA [31]. The temperature dependence of magnetization was measured by a superconducting quantum interference devise (SQUID) with an applied magnetic field of 10 Oe on MPMS-3 (Quantum Design). The temperature dependence of electrical resistivity was measured by a four-terminal method with an AC current of 1 mA on PPMS (Quantum Design). Au wires were attached on the sample surface by Ag paste. In this article, we have labeled the examined samples with the isotope mass (32 or 34) and alphabets (a–f) for clarity.

## 3. Results and discussion

We have synthesized polycrystalline samples at 540 °C because the reproducibility of the sample quality and the observed superconducting properties had been confirmed from experiments with natural sulfur chemical, while the obtained sample contained small amount of Bi$_2$S$_3$ and Bi impurities as indicated by + and * symbols as displayed in Fig. 1(a). Similar impurities were observed in previous reports on Bi$_4$O$_4$S$_3$ [13,27,29]. Since the crystal structure is complicated, as shown in Fig. 1(b), and there is possibility of stacking fault and modification of the stacking in the ternary Bi-O-S system [32-34], we have tried to synthesize Bi$_4$O$_4$S$_3$ isotope samples with close lattice constants to investigate the essential behavior on the isotope replacement between $^{32}$S and $^{34}$S. The lattice constants for the selected six samples are listed in Table I.

Figure 2 shows the temperature dependences of magnetization for Bi$_4$O$_4$S$_3$ samples with (a) $^{32}$S and (b) $^{34}$S. From the figure, the transition temperature in the magnetization data ($T_c^M$) looks almost the same. To evaluate $T_c^M$, temperature differential of magnetization ($dM/dT$) was calculated and plotted in Fig. 3. $T_c^M$ was defined as a temperature where two linear fitting lines for the data just below and above the transition as shown in Fig. 3. The estimated $T_c^M$ ranges from 4.73 to 4.76 K for $^{32}$S and $^{34}$S samples. Interestingly, there was no clear correlation between $T_c^M$ and the isotope mass in the magnetization data. The isotope exponent calculated from all the possible combination of $T_c^M$ is $-0.1 < \alpha_S < 0.1$.

Figure 4 shows the temperature dependences of electrical resistivity ($\rho$) for $Bi_4O_4S_3$ samples with (a) $^{32}S$ and (b) $^{34}S$. Typically, a superconducting transition onset for $Bi_4O_4S_3$ is quite broad and obviously higher than zero-resistivity temperature, which was explained as strong superconducting fluctuation [35]. Therefore, the onset temperature is not preferable to be used for the discussion of isotope effect. We hence estimated resistive $T_c$ ($T_c^\rho$) as a midpoint of the transition; the definition is displayed in Fig. 4(a). $T_c^\rho$ was defined as a temperature where $\rho(T)$ becomes 50% of $\rho(T = 5\ K)$. As listed in Table I, $T_c^\rho$ for $^{32}S$ samples ranges from 4.53 to 4.66 K, and $T_c^\rho$ for $^{34}S$ samples ranges from 4.54 to 4.64 K. Although the $T_c^\rho$ is relatively scattered as compared to the case of $T_c^M$, we could confirm that there is no clear correlation between $T_c^\rho$ and the isotope mass.

As mentioned above, the Bi-O-S phases have various structural forms including stacking fault [32-34]. Here, to avoid the possibility of the shift in $T_c$ due to those structural differences, we compare two ($^{32}S$ and $^{34}S$) samples having similar lattice constants. Accordingly, from table I, the set of #32-c and #34-e were selected. For those two samples, the $T_c^M$s are the same ($T_c^M = 4.73\ K$), and the $T_c^\rho$s are 4.53 K for #32-c and 4.64 K for #34-e. If the conventional phonon-mediated mechanisms were essential in a superconductor with a $T_c = 4.7\ K$, a difference in $T_c$ of 0.14 K is expected between $^{32}S$ and $^{34}S$ isotopes. Therefore, the isotope effect seen from the comparison between #32-c and #34-e clearly deviates from the expectation from the conventional phonon-mediated superconductivity mechanisms. The present results show that the unconventional isotope effect is essential in $Bi_4O_4S_3$ with $^{32}S$ and $^{34}S$.

Here, we briefly discuss about the commonality of the unconventional isotope effect in $BiS_2$-based superconductors. As mentioned in above, a similar unconventional isotope effect was observed in $LaO_{0.6}F_{0.4}BiSSe$ with BiSSe-type superconducting layers [26]. Although the constituent element of the superconducting layer is different, the in-plane structural symmetry is common because both are crystallized in a tetragonal structure ($P4/nmm$ for $LaO_{0.6}F_{0.4}BiSSe$ and $I4/mmm$ for $Bi_4O_4S_3$). In addition, $T_c$ is close; the typical $T_c$ of $LaO_{0.6}F_{0.4}BiSSe$ is 3.8 K. As well known, the $BiS_2$-based compounds exhibit a pressure-induced structural transition, at which a jump (abrupt increase) in $T_c$ is observed [36-37]. In the case of $LaO_{0.5}F_{0.5}BiS_2$, $T_c$ increases from 2 K (tetragonal $P4/nmm$) to 10 K (monoclinic ($P2_1/m$) by applying a pressure of 1 GPa [36]. Therefore, we can regard $LaO_{0.6}F_{0.4}BiSSe$ and $Bi_4O_4S_3$ with a tetragonal structure as a low-$T_c$ phase of the $BiS_2$-based superconductor from a viewpoint of in-plane structural symmetry, which is also a commonality between those two superconductors. To conclude what factor is essential for the emergence of unconventional pairing states in the $BiS_2$-based superconductors, isotope effect and other studies for a high-$T_c$ phase with a monoclinic phase is needed, and that is underway.

## 4. Conclusion

We have reported the sulfur isotope effect on $T_c$ in $Bi_4O_4S_3$ with $^{32}S$ and $^{34}S$. Polycrystalline samples of $Bi_4O_4S_3$ were prepared by solid-state reaction using $^{32}S$ and $^{34}S$ isotope chemicals. To investigate the isotope effect, six samples having similar lattice constants were selected. From magnetization analyses, the isotope exponent was estimated as $-0.1 < \alpha_S < 0.1$. Although the $T_c$ estimated from electrical resistivity was scattered as compared to those estimated from the magnetization, we observed no clear correlation between $T_c$ and the isotope mass. The present results on the isotope effect suggest $Bi_4O_4S_3$ to be an unconventional superconductor.

**Acknowledgements**

The authors thank O. Miura and Y. Goto for experimental supports. This work was partly supported by JSPS KAKENHI (Grant No. 18KK0076) and Advanced Research Program under the Human Resources Funds of Tokyo (Grant Number: H31-1).

**Table I. Sample label, lattice constants, and $T_c$ obtained from magnetization and resistivity ($T_c^M$ and $T_c^\rho$) for $Bi_4O_4S_3$ samples with $^{32}S$ and $^{34}S$.**

| Label | $a$ (Å) | $c$ (Å) | $T_c^M$ (K) | $T_c^\rho$ (K) |
|---|---|---|---|---|
| #32-a | 3.9667(4) | 41.337(5) | 4.76 | 4.57 |
| #32-b | 3.9723(4) | 41.367(6) | 4.74 | 4.66 |
| #32-c | 3.9710(3) | 41.368(4) | 4.73 | 4.53 |
| #34-d | 3.9694(5) | 41.358(6) | 4.74 | 4.59 |
| #34-e | 3.9708(3) | 41.361(4) | 4.73 | 4.64 |
| #34-f | 3.9732(4) | 41.384(4) | 4.76 | 4.54 |

**Figures**

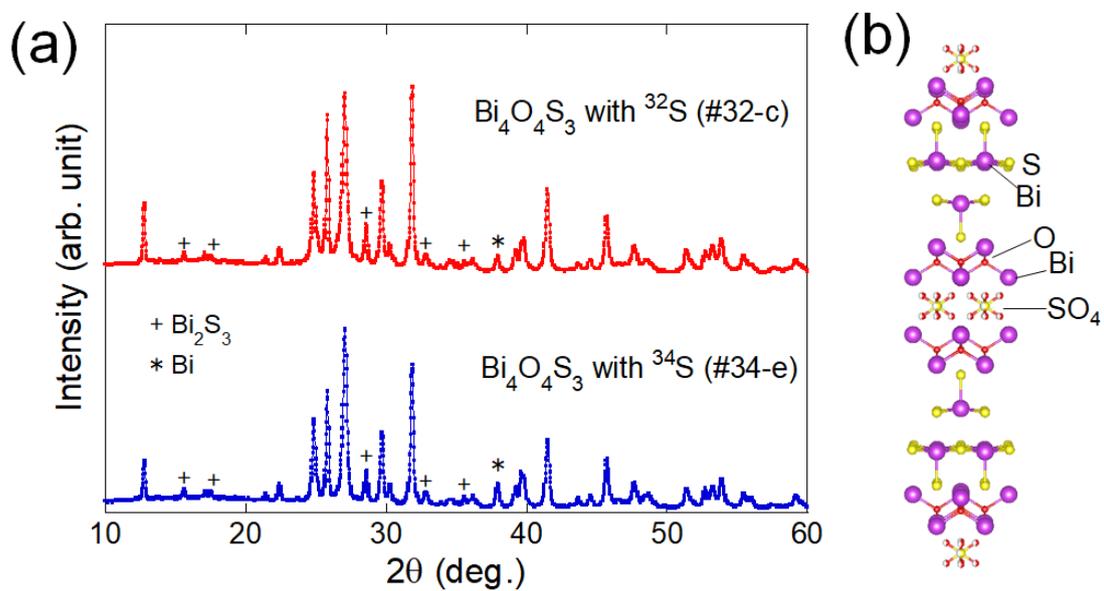

Fig. 1. (a) X-ray powder diffraction for Bi$_4$O$_4$S$_3$ samples prepared with $^{32}$S and $^{34}$S isotope chemicals. (b) Schematic image of crystal structure of Bi$_4$O$_4$S$_3$.

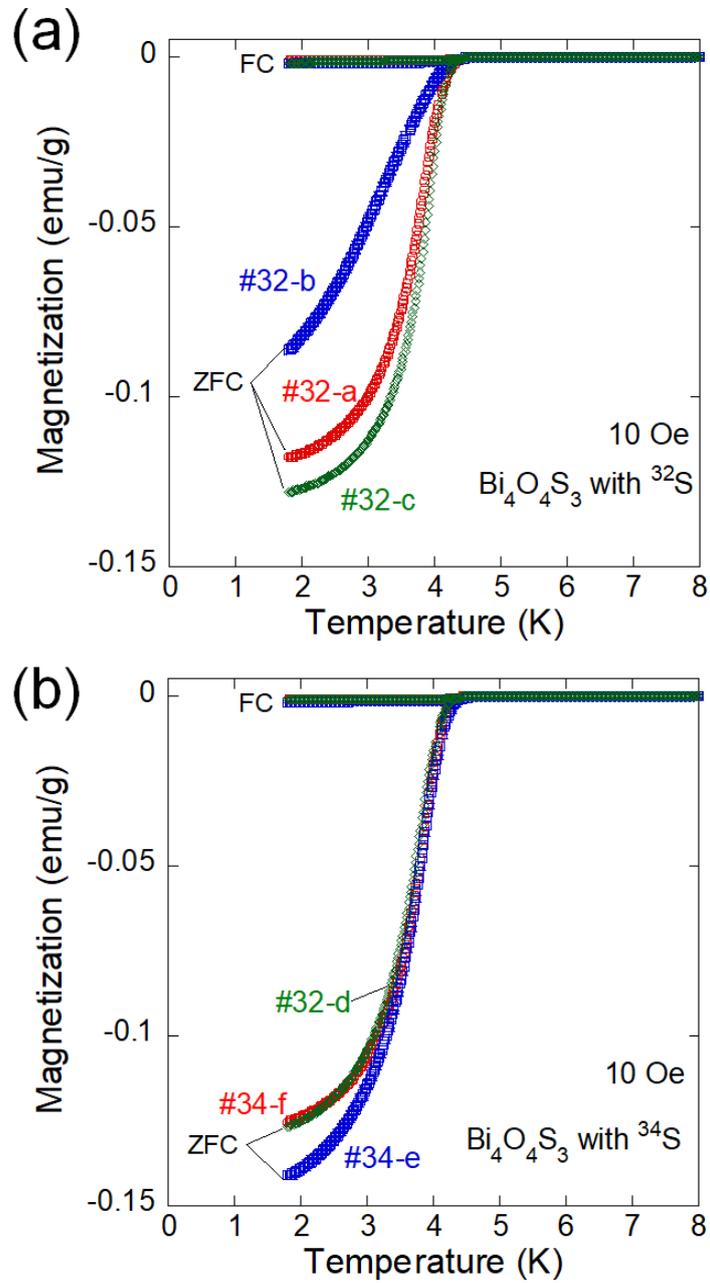

Fig. 2. Temperature dependences of magnetization after zero-filed cooling (ZFC) and field cooling (FC) for $Bi_4O_4S_3$ prepared with $^{32}S$ and $^{34}S$.

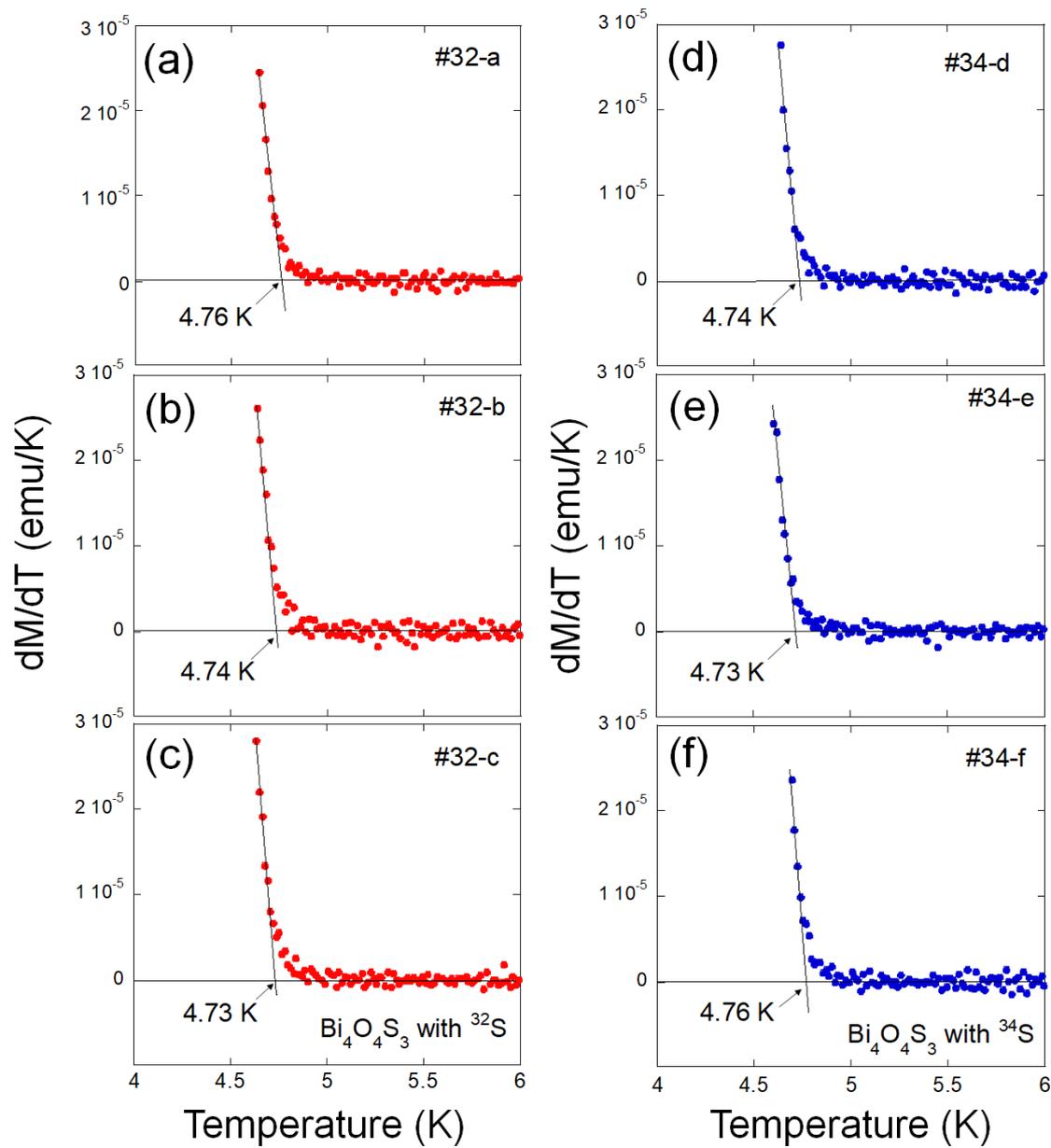

Fig. 3. Temperature differential of magnetization (ZFC) for $Bi_4O_4S_3$ prepared with $^{32}S$ and $^{34}S$.

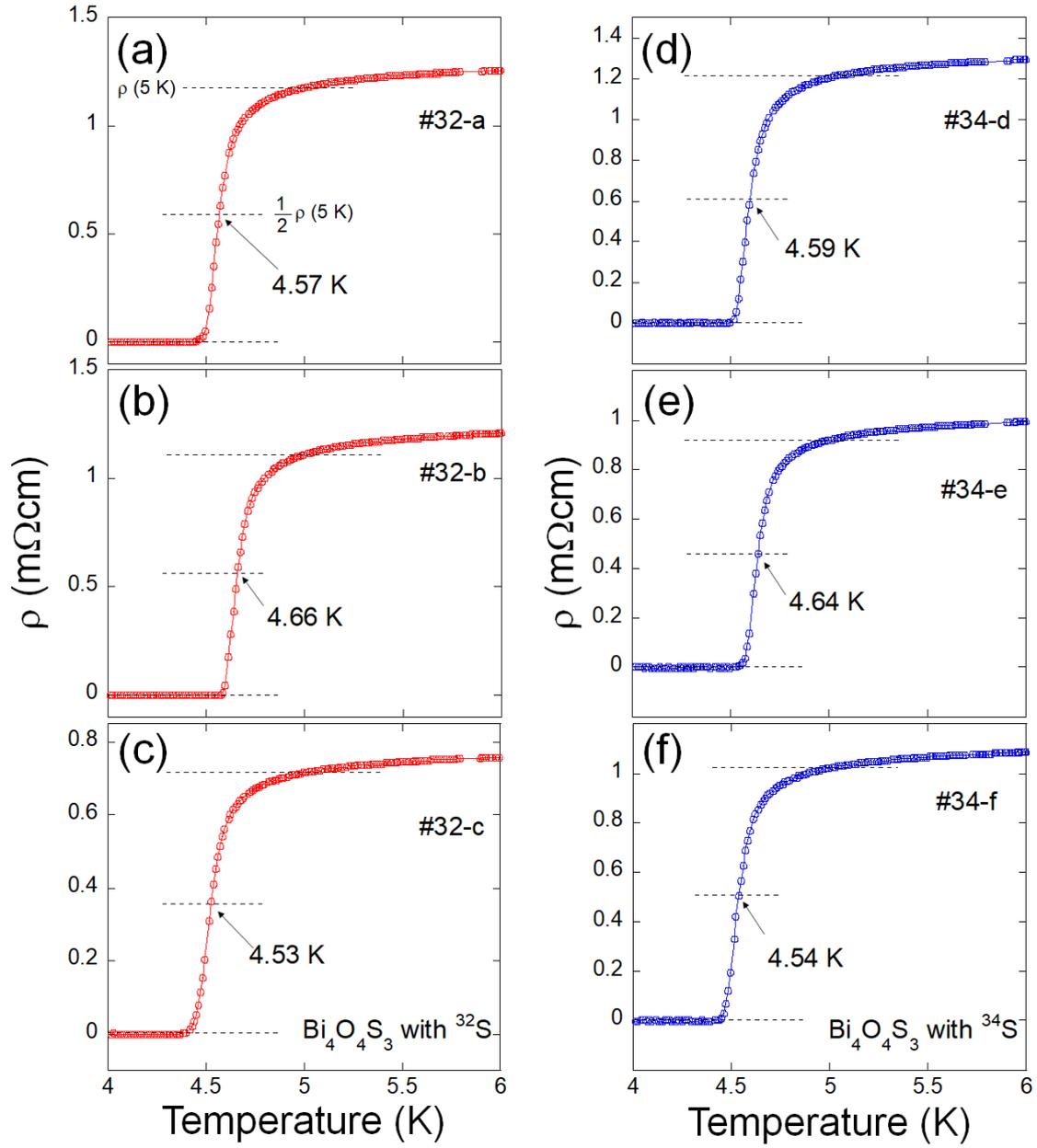

Fig. 4. Temperature dependences of resistivity ($\rho$) for $Bi_4O_4S_3$ prepared with $^{32}S$ and $^{34}S$.